\documentclass[aps,pra,floatfix,twocolumn,showpacs,superscriptaddress,reprint]{revtex4-1}
\usepackage{amsfonts}
\usepackage{mathrsfs}
\usepackage{amsmath}
\usepackage{color}
\usepackage{graphicx}
\usepackage{bm}
\usepackage{amssymb}
\usepackage{xspace}
\usepackage{dcolumn}
\usepackage{longtable}
\usepackage{multirow}
\usepackage{todonotes}


\begin{document}
\title{
 Measurement of collective excitations in a spin-orbit coupled Bose-Einstein condensate
       }

\author{M. A. Khamehchi}
\affiliation{Department of Physics and Astronomy, Washington State University, Pullman,  WA 99164, USA}
\author{Yongping Zhang}
\affiliation{Quantum Systems Unit, OIST Graduate University, Onna, Okinawa 904-0495, Japan}
\author{Chris Hamner}
\affiliation{Department of Physics and Astronomy, Washington State University, Pullman,  WA 99164, USA}
\author{Thomas Busch}
\affiliation{Quantum Systems Unit, OIST Graduate University, Onna, Okinawa 904-0495, Japan}
\author{Peter Engels}
\email{engels@wsu.edu}
\affiliation{Department of Physics and Astronomy, Washington State University, Pullman,  WA 99164, USA}

\begin{abstract}
We measure the collective excitation spectrum of a spin-orbit coupled Bose-Einstein condensate using Bragg spectroscopy. The spin-orbit coupling is generated by Raman dressing of atomic hyperfine states. When the Raman detuning is reduced, mode softening at a finite momentum is revealed, which provides insight towards a supersolid-like phase transition. We find that for the parameters of our system, this softening stops at a finite excitation gap and is symmetric under a sign change of the Raman detuning. Finally, using a moving barrier that is swept through the BEC, we also show the effect of the collective excitation on  the fluid dynamics.

\end{abstract}


\maketitle

Since the achievement of Bose-Einstein condensation (BEC) in dilute atomic gases, the investigation of collective excitations has been a key tool to gain insight into this unusual state of matter \cite{Ozeri}.
For most atomic species used in BEC experiments the interactions between the ultracold atoms can be described by isotropic, short-range s-wave scattering, which leads to the well-known linear phonon excitation spectrum at low momentum. However, if long-range interactions, such as dipolar interactions, are present, the collective excitation spectrum of a BEC can exhibit a more complex structure: in addition to the typical low energy phonon spectrum, a roton-like structure can appear. It is characterised by a shoulder in the spectrum, which for certain parameters can turn into a parabolic minimum at a finite momentum  \cite{Santos, Dell, Blakie,Jona}.

Interestingly, a similar parabolic minimum at a finite momentum can also exist in spin-orbit coupled (SOC) systems.  In cold atomic gases,  spin-orbit coupling can be implemented by Raman dressing of two or more atomic hyperfine states, which play the role of different (pseudo-)spins. The Raman lasers are arranged in such a way that a Raman transition between the states is accompanied by a change of momentum \cite{Lin, Fu, Zhang, Qu, Hamner,Olson}. Since the Raman coupling strength and the detuning from the Raman resonance can be independently adjusted in an experiment, this provides a very flexible platform to engineer interesting dispersion relations and test spin-orbit coupled physics (for a review, see, e.g., \cite{Dalibard, Galitski, Goldman, Zhai}).  In the single-particle picture, the effect of the Raman dressing is to displace two copies of the parabolic dispersion originating from the kinetic energy of the particle in opposite directions in momentum space. The Raman coupling opens a gap at the crossing of these two parabolas so that the resulting single-particle dispersion has the form of a double well in momentum space.  The double well can be biased towards either minimum by changing the Raman detuning. When the Raman detuning or the Raman coupling exceed a critical value, one of the minima disappears and a single well dispersion results.

\begin{figure}[t]
\scalebox{0.75}{\includegraphics*{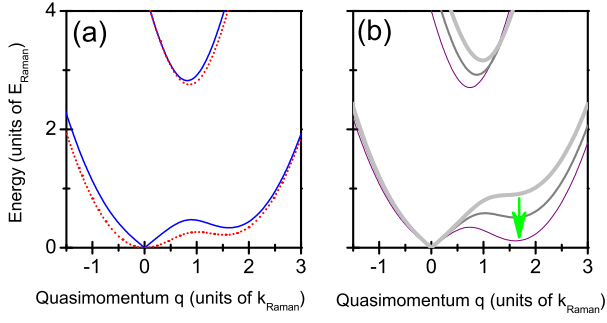}}
\caption{\label{BdGvsFree} (Color online) (a) BdG spectrum (blue solid line) and single-particle dispersion (red dotted line)  of a spin-orbit coupled BEC for a nonlinear coefficient of $g=0.186$, a Raman detuning of $0.28 E_\text{Raman}$ ($\delta = 2\pi \times 500$~Hz) and a Raman coupling strength of $2.5 E_\text{Raman}$. (b) Schematic of mode softening with decreasing Raman detuning. The lines correspond to a Raman detuning of $1 E_\text{Raman}$, $0.5 E_\text{Raman}$ and $0$ from top to bottom. }
\end{figure}

In the presence of nonlinear effects stemming from the s-wave scattering between the atoms in a Raman dressed BEC, the double well structure continues to exist. In a biased double well such as the one shown in Fig.~\ref{BdGvsFree} a BEC in the ground state occupies only one of the two double well minima \cite{Higbie}.  Compared to the single-particle spectrum, the second minimum is modified due to the interaction energy, and for small excitation momenta the collective excitation spectrum becomes linear (phonon modes). The resulting dispersion is reminiscent of the phonon-maxon-roton like structure \cite{Higbie, Martone, Zheng} that appears in the excitation spectrum of many interesting systems, such as  liquid $^4$He  \cite{Landau}, BECs with dipolar interactions  \cite{Santos, Dell}, fractional quantum Hall insulators \cite{Girvin} and antiferromagnets \cite{Oh}.

 Moreover, the presence of the nonlinearity leads to several phenomena within the collective excitation spectrum which are absent in the single-particle dispersion. One of the most fundamental of these is the softening of the roton-like modes when the spin-orbit coupling parameters are changed appropriately. This can trigger a first-order liquid-solid like phase transition when the gap below the roton minimum closes. Before reaching this phase transition point, a BEC in the ground state occupies only one of the two double-well minima, whereas after the phase transition it exists in a superposition of components located at both minima. This results in a crystalline structure in the spatial domain, which is a novel phase known as a one-dimensional supersolid possessing crystalline and superfluid properties at the same time  \cite{Li}.  In superfluids with long-range or finite-range interactions, roton softening provides a potential route to the formation of supersolids \cite{Kirzhnits, Pomeau1}.  Very recently, the roton minimum softening and a possible supersolid phase were observed experimentally for the first time in a BEC with cavity-mediated long-range interactions \cite{Mottl}.

Here we report on the experimental measurement of collective excitations in a spin-orbit coupled BEC by performing Bragg spectroscopy.   Previous theoretical studies have described roton-like excitations of SOC BECs for the case of vanishing Raman detuning \cite{Martone, Zheng}. In our experiment, we measure  mode softening  when, starting from a finite value, the Raman detuning is decreased. In the case of $^{87}$Rb, which is used in our experiment, the specific set of scattering lengths  would require a very low Raman coupling strength to reach the supersolid phase transition, which currently makes it unfeasible to collect measurable signals. For the experimentally used Raman coupling strength, the mode softening stops at a finite gap which protects the ground state against the phase transition. This experimental observation is corroborated by our theoretical analysis. Furthermore, we find a symmetry in the data for the mode softening which can be explained by a time-reversal like symmetry in the Gross-Pitaevskii Hamiltonian governing the system. Finally, we report on an experiment which might allow to study the influence of the roton-like mode on the hydrodynamic properties of the BEC.

%
%
%
%

The Raman dressing scheme in our experiment is based on coupling two hyperfine states along the $x$-direction. The two states play the role of (pseudo-)spins so that the BEC resembles an effective spin-$1/2$ system with spin-orbit coupled Hamiltonian \cite{Lin, Fu, Zhang, Qu, Hamner, Olson}
\begin{equation}
H_\text{soc}=-\frac{1}{2} \frac{\partial^2}{\partial x^2}  -i  \frac{\partial }{\partial x} \sigma_z +   \frac{\delta}{2} \sigma_z +\frac{\Omega}{2} \sigma_x,
\label{soc}
\end{equation}
where we have used $2 E_\text{Raman}=(\hbar k_\text{Raman})^2/m$ as the unit of energy, with
$2 \hbar k_\text{Raman}$ being the momentum imparted on the atoms during a Raman transition. The spin-orbit coupling  is characterized by  the term $  \partial /\partial x  \sigma_z$. The Raman detuning is denoted by  $\delta$  and the Raman coupling strength by $\Omega$.

For the experiments described below, the Raman dressing is adiabatically increased starting at large detuning and ending at a final detuning close to resonance. In this way the BEC is loaded near the minimum of the lowest spin-orbit band. Theoretically the BEC is described by the ground state of
the  mean-field Gross-Pitaevskii (GP) equation
\begin{equation}
 H_\text{soc} \Psi+ g \left ( |\Psi_1|^2 + |\Psi_2|^2  \right )  \Psi = \mu  \Psi,
\label{meanfield}
\end{equation}
where $\Psi=\left ( \Psi_1, \Psi_2 \right )^T$ is the spinor describing the two-components, $\mu$ is the chemical potential, and   $g=4\pi a k_\text{Raman}\text{N}$ with $a$ being the s-wave scattering length and $\text{N}$ being the atom number. We  assume that all scattering lengths are equal, which is a very good approximation for our experimental system \cite{scatteringlengths}.   While the theoretical analysis neglects the trapping potential present in the experiment, we will show below how this can be compensated for.  It is worth noting that the GP equations and the Hamiltonian $H_\text{soc}$ possess a  time-reversal-like symmetry described by $\mathcal{R}_{\delta} K \sigma_x$, where $\mathcal{R}_{\delta}$ flips the sign of the detuning,  $\mathcal{R}_{\delta} \delta  \mathcal{R}_{\delta}^\dagger =-\delta $, and $K$ is the operator for complex conjugation. This guarantees the symmetry of ground states of the GP equations when the sign of the detuning is changed.  It is also interesting to notice that a similar ground state of the spin-orbit coupled GP equations has been employed in Ref.~\cite{Williams} to produce higher-order partial waves in atomic collisions due to the long-range characteristic of the ground state.

Bragg spectroscopy is a powerful tool to investigate collective excitations in a BEC \cite{Stenger,Stamper}.
The linear response to a sudden perturbation through the Bragg pulse can be theoretically analyzed by the Bogoliubov-de Gennes (BdG) equations.  The general wavefunction spinor including the ground state, $\Phi_{1,2}$, and the perturbations, $\delta \Phi_{1,2}$, can be written as $\Psi_{1,2}(x,t)=e^{-i\mu t+ikx} \left [ \Phi_{1,2}(x)+  \delta \Phi_{1,2}(x,t) \right]$,
where $\mu$ and  $k$ are the chemical potential and quasi-momentum of the ground state, respectively. The perturbations can be parameterised  as  $ \delta \Phi _{1,2}(x,t) = U_{1,2}(x)\exp(iqx-i\omega
t)+V_{1,2}^*(x) \exp(-iqx+i\omega^* t) $, where  $U$, $V$, $q$,  and $\omega$ are the two amplitudes, the quasi-momentum and the frequency of the perturbations, respectively. After substituting the general wavefunctions into the time-dependent GP equations and retaining only terms linear in the perturbations, we arrive at the BdG equations
\begin{widetext}
\begin{equation}
\begin{pmatrix}H_1(k)+
(q+k)&\frac{\Omega}{2}+g\Phi_{1}\Phi_{2}^*&g\Phi_{1}^2&g\Phi_{1}\Phi_{2}\\
\frac{\Omega}{2}+g\Phi_{1}^*\Phi_{2}&H_2 (k)-
(q+k)&g\Phi_{1}\Phi_{2}&g\Phi_{2}^2\\
-g\Phi_{1}^{*2}&-g\Phi_{1}^*\Phi_{2}^*&- H_1 (-k)+
(q-k) &-(\frac{\Omega}{2}+g\Phi_{1}^*\Phi_{2})\\
-g\Phi_{1}^*\Phi_{2}^*&-g\Phi_{2}^{*2}&-(\frac{\Omega}{2}+g\Phi_{1}\Phi_{2}^*)&- H_2 (-k)-
(q-k)
\end{pmatrix}
\begin{pmatrix} U_1 \\ U_2 \\V_1 \\V_2\end{pmatrix} =\omega  \begin{pmatrix} U_1 \\ U_2 \\V_1 \\V_2\end{pmatrix},
\end{equation}
\end{widetext}
where
$H_1(k)=-\mu+\frac{(q+k)^2}{2}
+2g|\Phi_{1}|^2+g|\Phi_{2}|^2+\frac{\delta}{2}$ and $H_2(k)=-\mu+\frac{(q+k)^2}{2}
+g|\Phi_{1}|^2+2g|\Phi_{2}|^2 -\frac{\delta}{2}$.

\begin{figure}[t]
\scalebox{0.7}{\includegraphics*{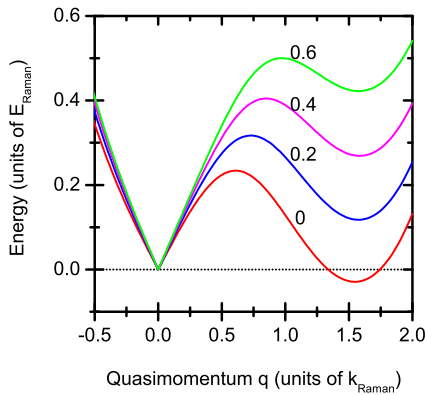}}
\caption{\label{Roton}(Color online) Roton-like minimum softening and energetic instability for decreasing  Raman detuning $\delta$ in a miscible regime ($g_{11}=g_{22}=0.186, g_{12}=0.08$) with
$\Omega=2.5 E_\text{Raman}$.  The lines correspond to the Raman detuning of $0.6 E_\text{Raman}$, $0.4 E_\text{Raman}$, $0.2 E_\text{Raman}$ and $0$ from top to bottom.}
\end{figure}

%
The resulting BdG spectrum and, for comparison, the single-particle dispersion of $H_\text{soc}$ for realistic parameters are shown in Fig.~\ref{BdGvsFree}(a).  The difference between the BdG and single-particle spectrum can clearly be seen.  The BdG spectrum is characterized by a phonon-maxon-roton-like feature with a local minimum near $1.62 \hbar k_\text{Raman}$ which is reminiscent of a roton minimum.  Reducing the Raman detuning allows to soften the roton-like mode, i.e. to decrease the excitation energy at the position of the minimum, without significantly affecting the long-wavelength phonon modes (Fig.~\ref{BdGvsFree}(b)). However, for practical parameters using  $^{87}$Rb, the energy of the minimum possesses a finite value even when the detuning vanishes. This is in contrast to the behavior of the single-particle spectrum, where the corresponding minimum goes to zero so that the two minima of the double well become degenerate. The finite gap in the BdG spectrum stabilizes the ground state and prevents a phase transition. We have checked numerically that in the phase miscible regime of scattering lengths, that allows a supersolid as a ground state \cite{Ho, Li2}, the softening of the roton-like minimum leads to a closure of the gap when a critical value of the Raman detuning is reached. The results are shown in Fig.~\ref{Roton} for a case of miscible parameters $g_{11}=g_{22}=0.186, g_{12}=0.08$. Below the critical value the roton-like modes possess negative energy, which indicates that a state occupying one of the two double-well minima is energetically unstable. The supersolid phase, which is a superposition of components at both minima, is then energetically preferred.

\begin{figure}[b]
\scalebox{0.3}{\includegraphics*[angle=90]{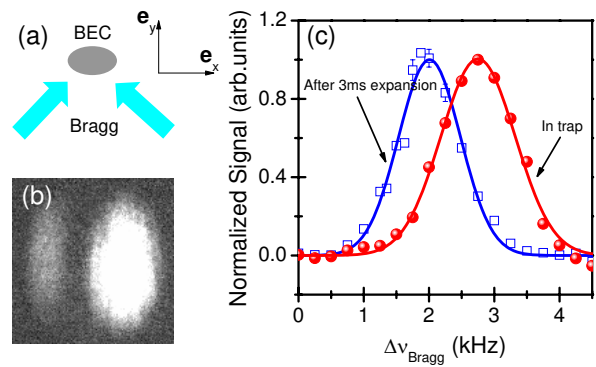}}
\caption{\label{BraggEnergyShift} (Color online) Bragg spectroscopy for BECs without spin-orbit coupling. (a) Schematic of the Bragg beam geometry. (b) Experimental image belonging to the blue data set (squared symbols) of (c), taken near the peak of the resonance. The fainter cloud on the left shows the Bragg scattered atoms. (c) Red dots: measured Bragg spectrum for a trapped BEC. Blue squares: spectrum obtained when the Bragg pulse is applied to a BEC after a 3~ms expansion time.  The blue data is an average over four measurements, typical error bars are shown near the peak. The red data is from a single measurement. The lines are Gaussian fits to the data.}
\end{figure}

To experimentally observe mode softening we perform Bragg spectroscopy on a $^{87}$Rb BEC. Our experiments start with a BEC confined in a crossed dipole trap with harmonic trap frequencies given by $\left(\omega_{x},\omega_{y},\omega_{z}\right)=2\pi\times\left(39,153,189\right)$~Hz.
Spin-orbit coupling in the $x$-direction is induced by two Raman laser beams with $\lambda_\text{Raman}\approx789$ nm, which intersect at the position of the BEC and are arranged with an angle of $90^{\circ}$ between each other. The Raman lasers couple the $|1,0\rangle=|1 \rangle$ and $|1,-1\rangle=|2\rangle$ hyperfine states in the $F=1$ manifold, which are split by a 10~G magnetic bias field. The accompanying quadratic Zeeman shift places the $|1,+1\rangle$ state $7.8~E_\text{Raman}$ away from resonance. For the Bragg spectrocopy, two Bragg laser beams with a wavelength of $\lambda\approx1540$~nm and small frequency difference, $\Delta\nu_\text{Bragg}$, are pulsed on for $1$~ms. The Bragg beams are collinear with the Raman beams (see Fig.~\ref{BraggEnergyShift}(a)) so that $k_\text{Bragg}=2\pi/(\sqrt{2}\cdot 1540$ nm). The BdG analysis presented above describes a homogeneous BEC, whereas the experimental system is confined in a harmonic trap. To remedy this discrepancy and account for the spatial variation of the density in the experiment, one can introduce an effective scattering length, $a_\text{eff}$. To determine the value of $a_\text{eff}$ for our experiment, we first measure the Bragg spectrum for a BEC with $10^5$ atoms without spin-orbit coupling (red dots in Fig.~\ref{BraggEnergyShift}(c)) and find a peak located at $\Delta\nu_\text{Bragg}$ = 2.7~kHz. This peak position can be reproduced by the formula for the BdG spectrum of a homogeneous BEC if the density is taken to be equal to
the central density of the trapped BEC and all scattering lengths are
set to $a_\text{eff} = 53.7a_0$ where $a_0$ is the Bohr radius \cite{scatteringlengths}. Due to the weakness of the trap in our experiment, it is a reasonable assumption that the same value of $a_\text{eff}$ is valid for the case of a BEC with spin-orbit coupling and we will show below that this leads to excellent agreement between theory and experiment. To demonstrate the dependence of the BdG spectrum on the interatomic interactions, we also show in  Fig.~\ref{BraggEnergyShift}(c) the spectrum for a BEC without the spin-orbit coupling that has been released from the trap and allowed to expand for 3~ms before the application of the Bragg pulse (blue squares). In this case, the interaction energy has predominantly been transformed into kinetic energy and the measured peak position, 1998~(17)~Hz, is close to the value expected for a single particle,  $E/h=\left( 2 \hbar k_\text{Bragg}\right)^{2}/2 m h = 1934$ Hz. This corresponds to the kinetic energy of an atom after receiving a momentum kick of $2 \hbar k_\text{Bragg}$ during the Bragg pulse.

\begin{figure}[b]
\scalebox{0.7}{\includegraphics*{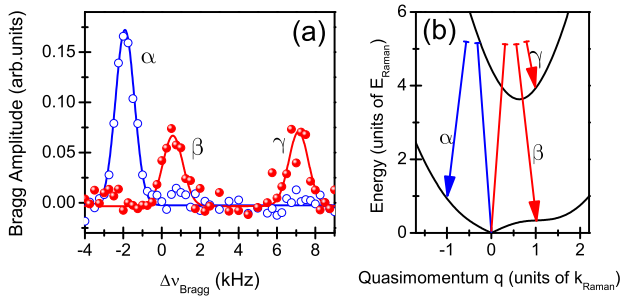}}
\caption{\label{peaks_500Hzdet} (Color online) (a) Bragg spectrum for a spin-orbit coupled BEC, measured for $\hbar \Omega=3.5 E_\text{Raman}$ and $\delta = 2\pi \times 500$~Hz. Each point is an average over four measurement. (b) Schematic of the transitions corresponding to the three peaks in the spectrum.}
\end{figure}

\begin{figure}[tb]
\scalebox{0.75}{\includegraphics*{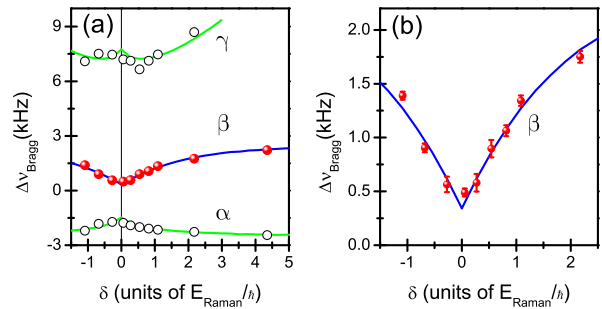}}
\caption{\label{RotonSoftening} (Color online) Mode softening. (a) Position of Bragg peaks as a function of Raman detuning. Each point is an average over four data runs.  This data was taken for $\hbar \Omega=3.5 E_\text{Raman}$. Vertical error bars in (a) are on the order of the symbol size. The data quality in the uppermost branch is impacted by the smallness of the spin overlap between the initial and final state. (b) Zoomed-in view of the data for peak $\beta$. The lines in (a) and (b) represent the result of theoretical calculations.}
\end{figure}

Next we apply Bragg spectroscopy to a BEC with spin-orbit coupling.  In order to observe the most interesting roton-like feature and its softening, the Bragg beam geometry, wavelength, and the spin-orbit coupling parameters should be chosen such that the momentum kick imparted by the Bragg pulse, when directed towards the roton, can transfer atoms to the roton minimum. For our geometry and Raman laser wavelength this leads to the choice of $\hbar \Omega=3.5 E_\text{Raman}$ for the data in Fig.~\ref{peaks_500Hzdet} and Fig.~\ref{RotonSoftening}, and the resulting roton feature is slightly less pronounced than the one shown in Fig.\ref{BdGvsFree}. Spin- and momentum resolved time-of-flight imaging subsequently allows to detect the number of atoms scattered by the Bragg pulse.
We find that each Bragg spectrum contains several distinct peaks. A typical spectrum for a BEC with $4 \times 10^4$ atoms, a Raman detuning of 500~Hz and a Raman coupling strength of $3.5 E_\text{Raman}$  is shown in Fig.~\ref{peaks_500Hzdet}(a). Each peak corresponds to a different Bragg resonance within the BdG band as indicated in  Fig.~\ref{peaks_500Hzdet}(b). The peak $\beta$ probes the region of the roton-like mode. The reduced amplitude of the peaks with positive $\Delta\nu_\text{Bragg}$ is due to the difference in spin composition of the initial and final states of the Bragg transition: unlike the Raman beams, the Bragg beams do not change the spin state due to their large detuning from the Rb D1 and D2 lines.

To demonstrate the mode softening, we measure Bragg spectra for a range of different Raman detunings, which effectively determine the relative importance of the mean-field interaction vs. spin-orbit effects, and record the positions of the Bragg peaks, see Fig.~\ref{RotonSoftening}. The three data sets shown correspond to the peaks $\alpha$ (lowest lying curve), $\beta$ (middle curve) and $\gamma$ (highest lying curve) of Fig.~\ref{peaks_500Hzdet}.  One can see that the peak positions significantly shift as a function of the Raman detuning, with the roton mode ($\beta$ peak, middle curve) clearly softening for a decreasing positive value.    When the Raman detuning becomes negative, the shape of the dispersion relation of the SOC BEC changes in such a way that the ground state of the BEC now has a quasimomentum opposite to the value at positive detuning. The symmetry between the data points for positive and negative detuning provides direct evidence for the existence of the time-reversal-like symmetry $\mathcal{R}_{\delta} K \sigma_x$. The fact that the energy of peak $\beta$ never reaches zero corresponds to the absence of a supersolid phase transition.

To demonstrate the excellent agreement between the data and the model using the homogeneous BdG equations with the effective interatomic scattering length of $a_\text{eff}=53.7a_0$ determined in the context of Fig.~\ref{BraggEnergyShift}, we overlay the data in Fig.~\ref{RotonSoftening} with the theoretically obtained curves. The effective scattering length leads to a nonlinear coefficient  in the GP  Eq.~\eqref{meanfield} of $g = 0.186$. The calculated spectrum is in very good agreement with the experimental data.

\begin{figure}[t]
\scalebox{0.75}{\includegraphics*{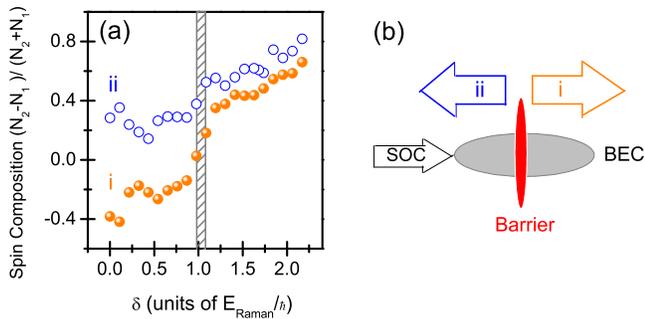}}
\caption{\label{barriersweep} (Color online) (a) Spin composition of a spin-orbit coupled BEC after a repulsive light sheet has been swept through it. The shaded rectangle indicates the region where the roton minimum starts to disappear. The roton does not exist to the right of this region.
The data was taken for $\hbar \Omega=2.5 E_\text{Raman}$, leading to very obvious roton-like structures for low Raman detunings as shown in Fig.\ref{BdGvsFree}. (b) Sweep  direction of the light sheet for the two data sets.  Drawing not to scale. }
\end{figure}

The existence of modes carrying roton-like structures can have direct consequences for the hydrodynamic behaviour of the BEC. As a particular example we probe the response of an SOC BEC to a repulsive light sheet that is swept through the condensate along the x-direction, see Fig.~\ref{barriersweep}. The light sheet is formed by a laser with a wavelength of 660~nm and Gaussian waist sizes of $w_x = 12 \mu$m and $w_{y} = 70 \mu$m. The central barrier height is approximately three times larger than the chemical potential of the BEC and the sweep velocity at $2.5$ mm/s exceeds the central speed of sound ($\approx 1$  mm/s). Note that a barrier sweep with these parameters leads to significant heating of the BEC. After the sweep we measure the spin composition $\frac{N_{2}-N_{1}}{N_{2} + N_{1}}$ of the cloud, where $N_{2}$ ($N_{1}$) is the number of atoms in the $|2\rangle$ ($|1\rangle$) state in the low-momentum components after time-of-flight imaging.
Two different scenarios are shown in Fig.~\ref{barriersweep}(b): a barrier moving towards the roton direction (orange filled circles), and a barrier moving in the opposite direction (blue open circles).
When the Raman detuning is chosen such that the BdG spectrum supports a roton minimum (i.e. to the left of the shaded region in Fig.~\ref{barriersweep}(a)), a significant difference in spin composition for the two cases is clearly visible. For  spin-orbit parameters that do not support a roton minimum (i.e. to the right of the shaded region), the difference in spin composition is much reduced. The fact that this change occurs around the region where the roton minimum disappears possibly indicates the excitation of roton quasiparticles by the moving light sheet. A more detailed investigation of these effects is the topic of future work.

We hope that the presented experimental results and their theoretical interpretation will stimulate further studies of the fluid dynamics in SOC BECs. In superfluid helium, the roton minimum limits the flow speed in the superfluid phase according to the Landau criterion. In our system, the Raman dressing, which is employed to generate the spin-orbit coupling, breaks Galilean invariance, and it would be interesting to test the Landau criterion in different moving frames. Furthermore, our experiment can potentially be extended to the case of a spin-orbit coupled lattice \cite{Hamner1} where rotons also exist but their manifestation is complicated \cite{Toniolo}.

{\bf Note added:}   During the completion of this manuscript, two further manuscripts appeared reporting the observation of roton excitations in  BECs \cite{Ha, Ji}.

This work was partly supported by Okinawa Institute of Science and Technology Graduate University. PE, MAK and CH acknowledge financial support from the National Science Foundation (NSF) through Grant No. PHY-1306662.

\bibliographystyle{apsrev4-1}

\end{document}